\documentclass[11pt,letterpaper]{article}
\usepackage[letterpaper, margin=2.5cm]{geometry}
\usepackage{amsmath,amsfonts,amsthm}
\usepackage{authblk}

\newcommand{\ee}{\mathop{}\!\mathrm{e}}
\numberwithin{equation}{section}
\newcommand{\R}{\mathbb{R}}
\newcommand{\N}{\mathbb{N}}

\newcommand{\C}{\mathbb{C}}
\theoremstyle{plain}

\newtheorem{theorem}{Theorem}[section]

\newtheorem{proposition}[theorem]{Proposition}

\theoremstyle{definition}

\usepackage{cancel}
\usepackage[T1]{fontenc}
\usepackage{mathtools}
\usepackage{booktabs}
\usepackage{colortbl}
\usepackage{amssymb}
\usepackage{cleveref}
\usepackage[shortlabels]{enumitem}
\setlist[enumerate,1]{label=\textnormal{(\emph{\roman*})}}
\usepackage{cases} 
\usepackage{cite}
\usepackage{array,multirow}
\definecolor{amber}{rgb}{1.0, 0.75, 0.0}
\definecolor{aogreen}{rgb}{0.0, 0.5, 0.0}
\definecolor{darksienna}{rgb}{0.24, 0.08, 0.08}
\allowdisplaybreaks
\newcommand{\dd}{\mathop{}\!\mathrm{d}}
\usepackage[pagewise]{lineno}
\newcommand*\patchAmsMathEnvironmentForLineno[1]{%
  \expandafter\let\csname old#1\expandafter\endcsname\csname #1\endcsname
  \expandafter\let\csname oldend#1\expandafter\endcsname\csname end#1\endcsname
  \renewenvironment{#1}%
  {\linenomath\csname old#1\endcsname}%
  {\csname oldend#1\endcsname\endlinenomath}}%
\newcommand*\patchBothAmsMathEnvironmentsForLineno[1]{%
  \patchAmsMathEnvironmentForLineno{#1}%
  \patchAmsMathEnvironmentForLineno{#1*}}%
\patchBothAmsMathEnvironmentsForLineno{equation}%
\patchBothAmsMathEnvironmentsForLineno{align}%
\patchBothAmsMathEnvironmentsForLineno{flalign}%
\patchBothAmsMathEnvironmentsForLineno{alignat}%
\patchBothAmsMathEnvironmentsForLineno{gather}%
\patchBothAmsMathEnvironmentsForLineno{multline}%
\begin{document}

\title{Behavior-induced oscillations in epidemic outbreaks with distributed memory:\\
beyond the linear chain trick using numerical methods}

\date{} 

\author[1,4]{Alessia And\`o}
\author[2,4]{Simone De Reggi}
\author[3,4]{Francesca Scarabel}
\author[1,4,*]{Rossana Vermiglio}
\author[5,6]{Jianhong Wu}
  
\affil[1]{Department of Mathematics, Computer Science and Physics, University of Udine, Via delle Scienze 206, 33100 Udine, Italy}
\affil[2]{Department of Mathematics, University of Trento, 
  Via Sommarive 14, 38123 Povo (Trento), Italy}  
\affil[3]{School of Mathematics, University of Leeds, Woodhouse, Leeds LS2 9JT, United Kingdom}  
\affil[4]{CDLab -- Computational Dynamics Laboratory, University of Udine, 33100 Udine, Italy}
\affil[5]{ Laboratory for Industrial and Applied Mathematics, Y-EMERGE, York University, Toronto M3J 1P3, Canada}
\affil[6]{Fields-CQAM Laboratory of Mathematics for Public Health, York University, Toronto M3J 1P3, Canada}

% corresponding author
\affil[*]{Correspondence: rossana.vermiglio@uniud.it.}

\maketitle

\begin{abstract}
We considered a model for an infectious disease outbreak, when the depletion of susceptible individuals is negligible, and assumed that individuals adapt their behavior according to the information they receive about new cases. In line with the information index approach, we supposed that individuals react to past information according to a memory kernel that is continuously distributed in the past.  We analyzed equilibria and their stability, with analytical results for selected cases.
Thanks to the recently developed pseudospectral approximation of delay equations, we studied numerically the long-term dynamics of the model for memory kernels defined by gamma distributions with a general non-integer shape parameter, extending the analysis beyond what is allowed by the linear chain trick. In agreement with previous studies, we showed that behavior adaptation alone can cause sustained waves of infections even in an outbreak scenario, and notably in the absence of other processes like demographic turnover, seasonality, or waning immunity. Our analysis gives a more general insight into how the period and peak of epidemic waves depend on the shape of the memory kernel and how the level of minimal contact impacts the stability of the behavior-induced positive equilibrium.
\end{abstract}

% \keywords{incidence-based social distancing; behavioral epidemiology; pseudospectral approximation; MatCont; periodic solutions; stability; infectious disease model; linear chain trick}

\section{Introduction}

Human behavior plays a crucial role in the spread of infectious diseases, as individual actions often change in response to available information. For instance, individuals may decide whether to get vaccinated on the basis of scientific evidence or circulating rumors, or, during an epidemic, they might reduce their social contacts or adopt protective measures (such as wearing face masks or washing hands) following media reports. As a result, over the past few decades, the \emph{behavioral epidemiology} of infectious diseases has attracted growing interest, and 
% as mathematical models increasingly tried to incorporate behavioral aspects to gain deeper understanding or more realistic predictions of the spread of diseases\cite{manfredi2013book}.
%.. there has been an increasing interest towards the \emph{behavioral epidemiology} of infectious diseases \cite{manfredi2013book}. 
%The focus is on mathematical models in which individuals can adapt their behavior, for instance by choosing whether or not to get vaccinated, or, if they are aware of an ongoing epidemic, by reducing their contacts or adopting protective measures (like using face mask or washing hands). 
% In the literature, 
numerous approaches have been proposed to incorporate behavioral feedback into mathematical models, drawing on concepts and methods from different disciplines, including economics and sociology. We refer to the recent reviews  \cite{BedsonIntegrated, FunkReview, WangStatistical} and to the monograph \cite{manfredi2013book} for a comprehensive overview on this field.

A common approach to model behavioral feedback is to assume that the transmission and/or vaccination rates depend on the present and/or past infections. One of the first examples dates back to 1978, when Capasso and Serio first proposed an SIR (Susceptibles, Infectious, Removed) model where the force of infection was assumed to be a nonlinear function of the current \emph{prevalence}, i.e., the total number of infected individuals~\cite{capasso1978generalization}.

Later, d'Onofrio, Manfredi, and colleagues considered models where information depends either on the past prevalence \cite{dOnofrioManfredi2009, dOnofrio2010vaccine, dOnofrioManfrediSalinelli2007vaccinating} or the past \emph{incidence}, i.e., the number of new cases per unit of time \cite{dOnofrioManfredi2022}. In this approach, typically known as the ``information index approach'', the delayed impact of infections on the current information %information at the present time 
is usually described by a continuous or concentrated memory kernel \cite{Ando2020fast}.
See also the recent works \cite{bulai2023stability,rossella,bulai2023covid}.

A common finding of these studies is that delayed behavioral feedback can cause sustained epidemic waves even in the absence of other processes that naturally induce oscillations, like seasonality, demographic turnover, and waning immunity. Even more complex dynamics emerge when additional factors are considered (e.g.,\cite{bulai2023stability,buonomo2023oscillations, sharbayta2022period}). 

Recently, Zhang and colleagues \cite{Zhang2023} considered an early-phase epidemic outbreak scenario, ignoring demographic turnover and depletion of susceptibles, and assumed that individuals adapt their behavior depending on the past incidence after a fixed delay $\tau>0$. This analysis was used to interpret the increasing peaks observed in the early multiple waves of diseases like SARS-1 and SARS-CoV-2. The model was later extended to an endemic scenario with loss of immunity \cite{cheng2025recurrent}.

In this paper, we generalize the model for disease transmission dynamics with the behavioral adaptation proposed in \cite{Zhang2023}. % Starting from a simple SIR model, we consider an early-stage outbreak scenario in which public health interventions, such as social distancing and personal protective measures, are promptly implemented to prevent a large proportion of infections before herd immunity is achieved and variants can emerge. 
We assume that the susceptible population remains constant, as both depletion of susceptibles and demographic turnover are negligible. This \emph{low attack rate} assumption, which is typically realistic at the onset of an infectious disease outbreak, may also hold over a longer time period. In fact, maintaining a low attack rate for as long as possible can be the only viable strategy against severe diseases in the absence of pharmaceutical interventions, as demonstrated by the recent COVID-19 pandemic~\cite{dOnofrioManfrediIannelli2021}.

The dynamics of the number of infected individuals, $I(t)$, are described by the classical linear ordinary differential equation (ODE)
\begin{equation} \label{I-ODE}
    I'(t) = \beta(t) I(t) - \gamma I(t),
\end{equation}
where $\beta$ is the per capita \emph{transmission rate}, given by the number of contacts per unit of time multiplied by the transmission probability per contact, and $\gamma$ is the per capita \emph{recovery rate}. 

To incorporate behavioral adaptation, we follow the implicit approach of \cite{dOnofrioManfrediSalinelli2007vaccinating,dOnofrioManfredi2009}. %and widely used since then.
We assume that 
\begin{equation}\label{beta}
    \beta(t) = f(M(t)),
\end{equation}
where $M(t),$ the \emph{information index},  captures a summary of the information available at time $t$, and $f$ is a given function. 
The information index depends on the past as \begin{equation}\label{iindex}
M(t)= \displaystyle \int_0^\infty G\left(t-s\right)K(s)\dd s, 
\end{equation} 
where $K$ is a \emph{memory kernel} and $G$ describes how individuals respond to the disease over time %through the \emph{message function} $g$, representing the information that they consider important when deciding whether or not to adopt precaution
\cite{buonomo2023oscillations, dOnofrioManfredi2009, dOnofrioManfredi2022, manfredi2013book}. 
In particular, $G$ can depend on time via the disease prevalence (e.g., if individuals react to the information about hospital occupancy, assumed proportional to the number of infected individuals), the incidence (e.g., if individuals react to the daily number of newly diagnosed cases), or deaths. 
% In the \emph{prevalence-based} case, the function $g$ is proportional to the infected individuals $I$, i.e. $g(s)=cI(s), \ c>0.$ It could be interpreted as the under-reporting coefficient of actual incidence, \cite{dOnofrioManfredi2022}, or the fraction of individuals who develop severe symptoms or will be hospitalized. In contrast a simple \emph{incidence-based} formulation assumes that $g$ is proportional to the incidence. 
In the \emph{prevalence-based} case, $G(t)=g(I(t)),$ and formula \eqref{beta} with \eqref{iindex} gives an explicit expression of $\beta$ as a function of the history of $I$, hence \eqref{I-ODE} becomes a delay differential equation with a distributed delay term. 
In this paper, we focus on the \emph{incidence-based} case, when information is driven by the number of new cases, represented in our model by the term $\beta I,$ so that $G(t)=g(\beta(t)I(t)).$ Therefore, the expression \eqref{iindex} defines $M$ as a function of the histories of both $I$ and $\beta$, and \eqref{beta} becomes a \emph{renewal equation} for~$\beta$.

Renewal equations are notoriously difficult to study. In addition, for general continuously distributed memory kernels, \eqref{beta} has infinite delay, and we refer to~\cite{DiekmannGyllenberg2012Blending} for the relevant dynamical systems theory and the principle of linearized stability.  To enable the mathematical analysis of behavioral epidemic models of this type, Erlang memory kernels are usually considered in the literature, as the resulting models can be reduced to a system of ODEs through the linear chain trick (LCT) \cite{Ando2020fast, cassidy2022numerical, dOnofrioManfredi2009}. 

Recently, a numerical technique has been introduced to reduce a renewal equation with infinite delay to an approximating system of ODEs. It is based on two main steps: first, the original equation is reformulated as an abstract differential equation on a suitable weighted Banach space; second, a system of ODEs is derived using the pseudospectral discretization approach \cite{Gyllenberg2018, Scarabel2024Infinite, IFAC2025}. 
One of the key advantages is that it allows us to analyze stability of equilibria and perform bifurcation analysis beyond Hopf bifurcations by available ODE software, like MatCont for MATLAB \cite{MatCont2008, Liessi-matcont}.
% while  a simple exposition of the treatment of the approximation of the iREs is given in \cite[Appendix A]{CISM-chapter-size} and \cite[Section 9]{CISM-Chapter-Age}. [potentially mention \cite{Liessi-matcont}] 

Crucially, the pseudospectral approximation technique can be applied to general (exponentially decaying) memory kernels, hence allowing the user to extend the analysis beyond Erlang distributions. We illustrate this strength by considering gamma-distributed kernels with general (continuous) shape parameters. 
Via numerical simulations, we investigate how the emergence of sustained epidemic waves depends on the parameters describing the memory kernel, and in particular on the average delay and the shape (determining the variance) of the distribution. 

%We should mention that an alternative numerical method has been recently proposed and applied to the integral formulation of models with infectivity depending on time since infection and including behavioral feedback \cite{buonomo2025minimal, buonomo2024stable, buonomo2025integral}.
%These methods allow to perform time integration of the model preserving the long-term properties of the system and, to our understanding, can also be applied to general memory kernels $K$. 

The paper is organized as follows. In Section~\ref{Smodel}, we introduce the incidence-based behavioral epidemic model for an emerging epidemic outbreak in the context of the information index theory. In Section~\ref{sec:stability}, we discuss the equilibria and their stability for general parameters, and present analytical results for some selected special cases. In Section~\ref{Snum}, we recall some essential details of the numerical approach, with a special focus on the numerical parameters involved. Section~\ref{Sresults} contains the numerical results that investigate the emergence of sustained oscillations when varying some model parameters. A discussion of the impact of our results in the public health context and some concluding remarks are given in Section~\ref{Sconclusion}.

\section{An infectious disease transmission dynamics model with behavioral adaptation and a continuously distributed information memory kernel}\label{Smodel}

In this section, we consider a model for an early-stage epidemic outbreak, which extends the one presented in \cite{Zhang2023} by incorporating distributed memory kernels.
We ignore births and deaths of individuals and assume that the host population is fully susceptible, i.e., $S(t)\equiv P$, where $S(t)$ denotes the number of susceptible individuals at time $t\ge 0$, and $P$ denotes the total population size.
Let $I(t)$ and $\beta(t)$ denote the number of infected individuals and the effective transmission rate, respectively, at time $t\ge 0$.
Assuming $I\ll P$, the \emph{incidence-based behavioral outbreak model} reads as follows: 
% Let $I(t)$ denote the number of infected individuals at time $t\ge 0$, and let $\beta(t)$ denote the effective contact rate at time $t\ge 0$.
% We consider an epidemic setting where we ignore births and deaths of individuals and we assume that the population is fully susceptible: $S(t)\equiv N$, where $N$ is the total population size (under the assumption $I\ll N$). 
% The \emph{incidence-based behavioral} model reads as follows:
\begin{equation}\label{model}
\left\{\setlength\arraycolsep{0.1em}
\begin{array}{rl} 
I'(t) &= \beta(t)I(t)-\gamma I(t),\\[2mm]
\beta(t)&= f\left(\displaystyle\int_0^\infty g\left(\beta(t-s)I(t-s)\right) K(s) \dd s\right),
\end{array} 
\right.
\end{equation}
where the message function $g\colon\R_{\ge 0}\to \R_{\ge 0}$ is increasing,  satisfies $g(0)=0$, and describes how the new infections  $\beta(t)I(t)$ at time $t$ ``translate'' into information, i.e., how strongly they reach public opinion (for instance, $g$ could account for under-reporting, or for the fact that the disease is not detected below a certain threshold); the memory kernel $K\colon \R_{\ge 0}\to \R_{\ge 0}$ satisfies
\begin{equation}\label{normK}
\int_0^\infty K(s)\dd s =1
\end{equation}
and describes how past infections affect the current ``news'' stream (this could be, for instance, captured by the time distribution from infection to diagnosis or death); $f\colon\R_{\ge 0}\to \R_{\ge 0}$ is decreasing and describes the effect of the current information and rumors on the transmission rate; and finally, $\gamma>0$ is the recovery rate from the disease.
% where we assume that
% \begin{itemize}[noitemsep]
% \item the \emph{message function} $g\colon\R_{\ge 0}\to \R_{\ge 0}$ is increasing with $g(0)=0$ and it describes how the new cases  $\beta(t)I(t)$ at time $t$ `translate' into information, i.e., how strongly they reach public opinion; $g$ could contain for instance an under-reporting factor, or the fact that the disease is not detected under a certain threshold; 
%     \item the \emph{memory kernel} $K\colon \R_{\ge 0}\to \R_{\ge 0}$ satisfies
% \begin{equation}\label{normK}
% \int_0^\infty K(s)\dd s =1.
% \end{equation} and it describes how past infections affect the current `news' stream (this could be, for instance, captured by the time distribution from infection to diagnosis or death);
%     \item $f\colon\R_{\ge 0}\to \R_{\ge 0}$ and it describes the effect of the current information and rumors on the transmission rate;
%     \item $\gamma>0$ is the \emph{recovery rate} from the disease.
% \end{itemize}
% For the function $f$, we assume \eqref{f1}

% where $k$ represents the \emph{plasticity/flexibility index} \cite{Zhang2023} \fra{or \emph{reactivity factor} \cite{buonomo2023oscillations}}: when the incidence is constant at the value $1/k$, the effective contact is half of the effective contact in the absence of infections: $1 + k\times \text{(daily incidence)} = 2$, and then the integral of $K$ is simply the average of a constant function.

Note that the model proposed in \cite{Zhang2023} can be recast as \eqref{model} by taking $K(s)=\delta(s-\tau)$, for $\tau>0$, where $\delta$ is the Dirac delta distribution concentrated at 0, and $g(x)= x$.
Furthermore, model \eqref{model} can be formulated in the framework of the \emph{information index approach} (see \cite{dOnofrioManfredi2009, dOnofrioManfredi2022}) via \eqref{beta} and \eqref{iindex}, with $G(t)=g(\beta(t)I(t))$. 
In particular, the paper \cite{dOnofrioManfredi2022} studies the emergence of oscillations for an incidence-based information index with exponentially fading and acquisition-fading kernels, such that the model can be reduced to a system of ODEs. An important difference between the models in \cite{dOnofrioManfredi2022} and \cite{Zhang2023} is that the former considers nonlinear dynamics with depletion of susceptibles (incidence term $\beta(t)S(t)I(t)$ where $S(t)$ is the number or fraction of susceptibles), while the latter focuses on an emerging disease outbreak situation where the number of immune individuals is assumed to be negligible (i.e., under a low attack rate assumption). 

In the following, we consider distributed memory kernels of the form 
\begin{equation}\label{gamma}
    K(s) = \cfrac{\mu^{\alpha}s^{\alpha-1}{\ee}^{-\mu s}}{\Gamma(\alpha)},\quad s \in \R_{\ge0},\quad \alpha, \mu >0,
\end{equation}
i.e., $K$ is a gamma distribution density function with shape parameter $\alpha$ and rate parameter $\mu$, so that the mean is $\tau:=\alpha/\mu$. As a result, \eqref{model} couples a linear ODE for $I(t)$ with a nonlinear renewal equation for $\beta$, as $\beta(t)$ is defined via the histories at time $t$ of both $I$ and $\beta$, defined respectively as the $\mathbb{R}$-valued functions $I_t(\theta):=I(t+\theta),\   \beta_t(\theta):=\beta(t+\theta)$, for $\theta \in \mathbb{R}_{\leq 0}$ .  For such coupled differential-renewal equations with infinite delay, the natural choice of history spaces, i.e., where the history functions $I_t$ and $\beta_t$ live, is given by weighted Banach spaces defined using the exponential weight
\begin{equation}\label{w}w(\theta)=\ee^{\rho\theta}, \quad \theta \in \mathbb{R}_{\leq 0},
\end{equation}
for $\rho>0$ (see, for example, \cite{DiekmannGyllenberg2012Blending}). In this setting,  $I_t$ is such that $w I_t$ is continuous and vanishes at minus infinity, while $\beta_t$ is absolutely integrable with respect to $w.$

%i.e. there exists $\rho>0$ such that for $w(\theta)=e^{\rho\theta},$ we have that $wI_t$ belongs to $L^\infty$ and $w(\theta)I_t(\theta) \atop  0$ as $\theta \to -\infty$ and $$w\beta_t \in L^1 $ 

\section{Equilibria and stability analysis}\label{sec:stability}

In this section, we compute the equilibria of system \eqref{model} and investigate, under appropriate assumptions, conditions for their stability. This analysis is carried out for general model parameters $K$, $f,$ and $g$, by applying the principle of linearized stability for infinite delay equations \cite{DiekmannGyllenberg2012Blending}. 

In particular, for a given equilibrium, the characteristic equation for the relevant linearized system is derived, and the local stability of the equilibrium is determined by the position in the complex plane of the (finitely many) characteristic roots in $\mathbb{C}_\rho=\{ \lambda \in \mathbb{C} \ | \ \Re \lambda >-\rho \},$ where $\rho$ is the parameter defining the weight \eqref{w}.

%Infinite delay equations generate infinite-dimensional dynamical systems in suitable exponentially weighted Banach space of functions. Let $w(\theta)=\ee^{-\rho \theta}, \theta \leq 0$, for $\rho >0,$ be the weight, and consider the linearized system at a steady state. Then a characteristic equation can be derived, whose roots with real part greater than $-\rho$ give the local stability of the steady-state. In particular if all the characteristic roots have negative real part, then the steady-state  is exponentially stable, and if there exists at least one characteristic root of  with positive real part, then the steady-state is unstable. 

Let us observe that $(I^*,\beta^*)\in \mathbb R^2$ is an equilibrium for \eqref{model} if and only if it satisfies
\begin{equation*}
\left\{\setlength\arraycolsep{0.1em}
\begin{array}{rcc} 
I^*\left(\cfrac{\beta^*}{\gamma}-1\right) &=&0,\\[4mm]
\displaystyle f \left(g\left(\beta^*I^*\right)\int_0^\infty K(s)\dd s \right)&=&\beta^*.
\end{array} 
\right.
\end{equation*}
Using \eqref{normK}, the second equation reduces to 
\begin{equation*}
\beta^* - f\left(g\left( \beta^* I^*\right) \right) = 0.
\end{equation*}
Let $\beta_0:=f(0)$. 
Since $g$ is increasing, $g(0)=0$, and $f$ is decreasing, it is not difficult to see that only two equilibria are possible, namely the disease-free equilibrium (DFE)
\begin{equation*}
E_0:=(0, \beta_0),
\end{equation*} 
and the established equilibrium (EE)\footnote{Here we use ``established equilibrium'' instead of the term ``endemic equilibrium'' that is normally used in epidemic modeling, as our focus here is on disease outbreak. We use established equilibrium to avoid confusion, as this is an equilibrium reached by the system due to behavior {adaptation} in an outbreak situation.} 
\begin{equation}\label{EEeq}
    E_+ :=\left(\frac{g^{-1}(f^{-1}(\gamma))}{\gamma} , \gamma \right).
\end{equation} 
In particular, $E_+$ exists and is positive if and only if $R_0 >1$,
where $R_0=\beta_0/\gamma$ is the basic reproduction number for \eqref{model}.
% Note that, since $g$ is nondecreasing and $f$ is nonincreasing, $E_+$ exists positive if and only if $R_0 >1$. 
Note that the equilibria are independent of the kernel $K$. 

To investigate the stability of equilibria, we first linearize \eqref{model} around $(I^*, \beta^*)$. From the first equation, we get
\begin{equation*}
I'(t) = I^* \beta(t) + \beta^* I(t) - \gamma I(t),
\end{equation*}
while from the second equation, we obtain 
\begin{align*}
    \beta(t) &= \displaystyle f'\left(\int_0^\infty  g\left(\beta^* I^* \right) K(s) \dd s  \right) \times \int_0^\infty g'\left(\beta^* I^* \right) \left(\beta(t-s)I^*  + \beta^* I(t-s) \right) K(s) \dd s \\
        &= \displaystyle C(I^*,\beta^*) \int_0^\infty \left(\beta(t-s) I^* + \beta^* I(t-s) \right) K(s)\dd s,
\end{align*}
where $C(I^*,\beta^*) := f'\left( g\left( \beta^* I^* \right) \right) \, g'\left( \beta^* I^* \right).$ Note that the assumptions on $f$ and $g$ ensure that $C(I^*,\beta^*)< 0.$
To apply \cite[Theorem 3.15]{DiekmannGyllenberg2012Blending}, we consider $\mu >\rho,$ and we derive the characteristic equation looking for solutions of the form 
\begin{equation*}
    \left(I(t), \beta(t)\right)=\left(\xi, \eta \right){\ee}^{\lambda t},\quad (\xi, \eta) \in \C^2 \setminus\{(0,0)\}\text{ and } \lambda \in \mathbb{C},\ \Re\lambda>-\rho.% ,\quad t\ge 0.
\end{equation*}
Let 
$$\widehat{K}(\lambda) := \int_0^\infty K(s) \,{\ee}^{-\lambda s}\dd s$$
denote the Laplace transform of $K$. From straightforward computations, we obtain the characteristic equation 
\begin{equation}\label{CE}
\Delta(\lambda)=0
\end{equation} 
where 
\begin{equation*}
\Delta(\lambda):=
\det\,\left(
\begin{array}{cc} 
\lambda +\gamma - \beta^*& -I^*\\
-C(I^*,\beta^*) \widehat{K}(\lambda) \beta^*&1-C(I^*,\beta^*) \widehat{K}(\lambda) I^*\\
\end{array}
\right).
\end{equation*}
%where $\hat{K}(\lambda)= \int_0^\infty \ee^{-\lambda s} K(s)\dd s$
%Hence, the linearization of \eqref{model} around the DFE $E_0$ reads 
%\begin{equation*}
%\left\{\setlength\arraycolsep{0.1em}
%\begin{array}{rl} 
%I'(t) &= (\beta_0-\gamma) I(t) \\[2mm]
%\beta(t) &= f'(0) g'(0) \beta_0\displaystyle %\int_0^\infty I(t-s)K(s)\dd s.
%    \end{array} 
%\right.
%\end{equation*}
%Now, we look for solutions of the form 
%\begin{equation}\label{expsol}
 %   \left(I(t), \beta(t)\right)=\left(\xi, \eta \right){\e}^{\lambda t},\quad \lambda, \xi, \eta \in \C,\quad t\ge 0,
%\end{equation}
%obtaining the characteristic equation 
%\begin{equation*}
%\left\{\setlength\arraycolsep{0.1em}
%\begin{array}{rl} 
%\lambda &= \beta_0-\gamma \\[2mm]
%    \eta &= \xi f'(0) g'(0) \beta_0 \displaystyle\int_0^\infty \ee^{-\lambda s} K(s)\dd %s. 
%\end{array} 
%\right.
%\end{equation*}
For the DFE $E_0$, \eqref{CE} reduces to $\lambda+\gamma-\beta_0=0,$ and we immediately get the following result.
\begin{proposition}
Let $\rho < \mu$. The DFE $E_0$ is locally asymptotically stable (LAS) when $R_0<1$ and unstable when $R_0>1$, regardless of the memory kernel $K$. 
\end{proposition}
%\begin{proof}$The claim follows from $f'(0) < 0$ and $g'(0)>0$. 
%\end{proof}
Now, we consider the EE. Let $\beta_+=\gamma,$ $I_+=\frac{g^{-1}(f^{-1}(\gamma))}{\gamma},$ and $C_+=C(I_+,\beta_+).$ 
%The linearization of model \eqref{model} around $E_+$ reads
%\begin{equation}\label{linmodel}
%\left\{\setlength\arraycolsep{0.1em}
%\begin{array}{rl} 
%I'(t) &= \bar I\beta(t),\\[2mm]
%\beta(t)&= \displaystyle f'\left( g\left(\bar\beta \bar I\right) %\right) \, g'\left(\bar\beta \bar I\right) \, \int_0^\infty %\left(\beta(t-s) \bar I+\bar \beta I(t-s)\right) K(s)\dd s.
%\end{array} 
%\right.
%\end{equation} 
%In order to recover a characteristic equation for \eqref{linmodel}, we look for solutions of the form \eqref{expsol}. 
%Then, inserting \eqref{expsol} into \eqref{linmodel}, and defining 
%\begin{equation*}
%$$    c_0:= f'\left( g(\bar\beta \bar I) \right) \, g'(\bar\beta %\bar I) < 0,
%\end{equation*}
%we obtain
%\begin{equation*}
%\left\{\setlength\arraycolsep{0.1em}
%\begin{array}{rl} 
%& \lambda \xi = \bar I \, \eta,\\[2mm]
%& \eta = \displaystyle c_0 \, \int_0^\infty \left( \eta \bar I+ %\bar \beta \xi \right) K(s) \ee^{-\lambda s} \dd s.
%\end{array} 
%\right.
%\end{equation*}
%From \eqrfe{CE}, we get the characteristic equation  %$\det\left(\Delta(\lambda)\right)=0$
%for the characteristic function
%\begin{equation*}
% \Delta(\lambda)=
% \begin{pmatrix}
%    \lambda & -\bar I \\
%- c_0 \bar \beta \displaystyle\int_0^\infty K(s){\e}^{-\lambda %s}\dd s & 1- c_0 \bar I \displaystyle\int_0^\infty K(s){\e}^{-%\lambda s}\dd s
% \end{pmatrix},\quad \lambda \in \C.
%\end{equation*}
%Let 
%$$\widehat{K}(\lambda) = \int_0^\infty K(s){\e}^{-\lambda s}\dd s,$$
%denote the Laplace transform of $K$. 
From \eqref{CE}, we get the following equation:
\begin{equation}\label{CEpiu}
    \lambda - C_+ I_+ \widehat{K}(\lambda) \left( \lambda + \beta_+ \right) = 0.
\end{equation}

Note that, differently from the case of the DFE, the stability properties of the EE depend on the particular choice of $K$. %, see, e.g.,  \cite{capasso1978generalization, dOnofrioManfredi2009, dOnofrioManfredi2022}.
For instance, in the \emph{memoryless} case, that is, when $K$ is a Dirac delta concentrated at $t=0$,  $\widehat{K}(\lambda)=1$ and the characteristic equation \eqref{CEpiu} reduces to 
\begin{equation*}
 \lambda (1-C_+  I_+) - C_+ \beta_+  I_+ = 0
 \quad \Leftrightarrow \quad 
 \lambda = \frac{C_+ \beta_+  I_+}{1-C_+  I_+} < 0.
\end{equation*}
Hence, the EE is always LAS when it exists, meaning that no sustained oscillations are possible when the behavioral response is not delayed, in accordance with  \cite{capasso1978generalization, dOnofrioManfredi2009, dOnofrioManfredi2022}.

Consider now the case of a fixed discrete delay between infections and consequent behavioral change, that is, we take $K$ as a Dirac delta concentrated at $\tau$. Then $\widehat{K}(\lambda)=\ee^{-\lambda\tau}$ and 
the characteristic equation \eqref{CEpiu} reads 
\begin{equation*}
    \lambda - C_+ I_+ \ee^{-\lambda\tau} \left( \lambda + \beta_+ \right) = 0.
\end{equation*}
This case is studied in \cite{Zhang2023}, where the authors
show that the EE could become unstable via a Hopf bifurcation when varying $\tau$.

For the gamma probability density function \eqref{gamma}, the Laplace transform of $K$ reads
\begin{equation}\label{KLaplace}
    \widehat{K}(\lambda) = \left( \frac{\mu}{\mu+\lambda} \right)^\alpha, \qquad \Re\lambda > -\rho > -\mu . 
\end{equation}
% Note that, when $\alpha\to \infty$ while $\alpha/\mu \to \tau < \infty$, the distribution tends to a Dirac delta in $\tau$.
Hence, for general shape $\alpha$ and rate $\mu$, the characteristic function reads as
\begin{equation}\label{quasipol}
    \Delta(\lambda)=\lambda (\mu+\lambda)^\alpha - C_+\mu^\alpha I_+ \lambda - C_+ \mu^\alpha \beta_+I_+. 
\end{equation}
If $\alpha$ is an integer (i.e., $K$ is Erlang), then $\Delta(\lambda)$ is a polynomial and one can try to study the stability of the EE through the Routh--Hurwitz criterion, see, for instance, \cite{dOnofrioManfredi2022}. 
We list here some possible choices for $\alpha \in \N$ that we will use later in the paper to validate the results obtained with the numerical method proposed in \cite{Scarabel2024Infinite}. Here we only illustrate the analytical results and refer to Appendix \ref{CompLinStab} for detailed computations. 
\begin{itemize}
\item Case $\alpha =1$: The EE is LAS for all $\mu>0$ 
(in agreement with what was observed in \cite{dOnofrioManfredi2022} for the model including depletion of susceptibles and demography).
\item Case $\alpha=2$: Applying the Routh--Hurwitz criterion, we obtain that the EE is LAS if 
\begin{equation*}
    \mu > \frac{-C_+ \beta_+I_+}{2(1-C_+ I_+)} > 0.
\end{equation*}
Hence, if $\mu$ is small enough (i.e., if the delay $\tau=2/\mu$ is large enough),  $E_+$ can be destabilized.
\item Case $\alpha=3$: Applying 
the Routh--Hurwitz criterion, we obtain that the EE is LAS if 
\begin{equation*}
\mu > \min \left\{ \frac{-C_+ \beta_+I_+}{1-C_+I_+}, \frac{-C_+\beta_+I_+}{1-C_+I_+} \cdot \frac{9}{8+C_+I_+} \right\}. 
% \left\{\setlength\arraycolsep{0.1em}
% \begin{array}{rl} 
% \mu &> \frac{-c_0 \bar{\beta}\bar{I}}{1-c_0\bar{I}}, \\[2mm]
% \left(\frac{8+c_0\bar{I}}{9} \right) \mu &> -\frac{c_0\bar{\beta}\bar{I}}{1-c_0\bar{I}}.
% \end{array} 
% \right.
\end{equation*}
\end{itemize}
For non-integer values of $\alpha$, it is hard to conclude anything about the roots of the characteristic equations. 
%, we cannot say anything on the roots of the characteristic equation. 
To overcome this issue, in the following, we will rely on numerics.

\section{Pseudospectral approximation by means of a system of ODEs} \label{Snum}

As discussed in the previous section, analytical stability results can be  derived  only for some special kernels. For more general cases, to investigate the stability of equilibria, one can rely on numerical methods. 

A recently proposed numerical method to approximate the stability and bifurcations of equilibria of nonlinear infinite delay equations is based on the \emph{pseudospectral}, or \emph{spectral collocation}, method. We here summarize the core ideas of this approach before applying them to some test problems in the next section. We refer to \cite{Gyllenberg2018, Scarabel2024Infinite} and the references therein for further details and rigorous error bounds, and to \cite[Appendix A]{DiekmannScarabelSize} for a concise and less technical summary.  
%versione uno
The coupled differential-renewal equation \eqref{model} can be reformulated as a semilinear abstract differential equation for the state $u(t):=(wI(t),w I_t,w\int_0^{\cdot}\beta_t) %\in L^\infty(\mathbb{R}_{\leq 0},\mathbb{R}^2) \times AC(\mathbb{R}_{\leq 0},\mathbb{R})
,$ with weight \eqref{w}, as follows: 
\begin{equation}\label{AbsEq}
%\frac{\dd}{\dd t}
u'(t)=A_0u(t)+F(u(t)), 
\end{equation}
where 
%\begin{equation}\label{A0}
% \left\{\setlength\arraycolsep{0.1em}
% \begin{array}{l} 
\begin{equation}\label{A01}
A_0(wa,w\psi,w\phi):=(0,w\psi',w\phi') 
\end{equation}
is linear with domain
\begin{align}\label{A02}
\notag D(A_0):= \{(wa,w\psi,w\phi)\ \mid\   & w\psi,\ w\psi'\in L^\infty(\mathbb{R}_{\leq 0},\mathbb{R}),\\[0.5mm] & \psi(0)=a\in\mathbb{R}, \text{ and } w\phi,\ w\phi'\in AC(\mathbb{R}_{\leq 0},\mathbb{R}),\ \phi(0)=0 \},
 %\\\end{array} \right.\end{equation}
 \end{align}
and the nonlinear part is given by
\begin{equation}\label{F}
F((wa,w\psi,w\phi)):=\left(wa \,(\phi'(0)-\gamma),\ 0,\ -wf\left(\int_0^\infty g(\psi(-s)\phi'(-s))K(s)ds\right)\right).
\end{equation}
The main idea of the spectral collocation approach is to approximate the solution of a differential equation by an interpolating function that satisfies the equation at a finite set of collocation points. In our case, we approximate both the differential and renewal components of the state by using weighted polynomials with domain $\mathbb{R}_{\leq 0}$ and fixed degree $N \in \mathbb{N}.$  Given the collocation nodes $\theta_i \in \mathbb{R}_{\leq 0}$, $i=1,\dots,N,$ by requiring %to 
the weighted polynomials to satisfy the collocation equations and domain conditions specified at $\theta_0=0$ in \eqref{A01}\textendash\eqref{A02}, we reduce the abstract equation \eqref{AbsEq} to the system of $2N+1$ ODEs% of dimension $2N+1$ 
\begin{equation*}
%\frac{\dd}{\dd t}
U'(t)=A_{0,N}U(t)+F_N(U(t)), 
\end{equation*}
where $F_N$ can also account for the use of a quadrature rule to approximate the integral appearing in \eqref{F}. In particular the delay differential component $(wI(t),w I_t)$ is approximated by the first $N+1$ differential equations, i.e., $U_i(t) \approx w(\theta_i)I(t+\theta_i),$ $i=0,1,\dots,N,$ while the renewal component $w\int_0^{ \; \cdot }\beta_t$ is approximated by the last $N$ differential equations, i.e., $U_{N+i}(t) \approx w(\theta_i)\int_0^{\theta_i}\beta(t+s)\dd s,$ $i=1,\dots,N$ \cite{Scarabel2024Infinite}.

An important strength of this approach is that the resulting approximating system can be analyzed with software for numerical continuation and bifurcation of ODEs, which are widely available, generally well developed, and well maintained. This offers an effective alternative to numerical simulations, and enables us to explore the dynamics of \eqref{model} with a gamma-distributed memory kernel, as shown in the next section.

In addition to the collocation degree $N$, the discretization requires us to select the parameter $\rho$ defining the weight \eqref{w}, a set of collocation nodes, and a quadrature rule to approximate integrals on $\mathbb{R}_{\leq 0}$ for the trade off between accuracy and computation time.
While the parameter $\rho$ is related to the history space of the original equation itself, both $N$ and $\rho$ play a direct role in the definition of the collocation nodes and quadrature rules---which are taken as the extrema of the classical Laguerre orthogonal polynomials scaled by $1/(2\rho)$ and the corresponding Gauss--Radau quadrature rule, respectively. The numerical approximation can be computed with numerically stable algorithms for weighted polynomial interpolation and weighted pseudospectral differentiation \cite{Mastroianni, weideman2000}. 

Finally, to improve computation times, truncated interpolation and quadrature rules can be used to obtain approximations with polynomials of the same degree $N$, but using an approximating system of lower dimension $N_r<N$, where $r \in (0,1)$ is an arbitrary truncation parameter. Given $r$, the reduced approximating system is constructed by considering only the $N_r$ variables corresponding to the $N_r$ nodes with modulus smaller than $2rN/\rho$. The truncation is also useful to avoid cancellation errors in the computation of the smallest quadrature weights. We refer to \cite{IFAC2025} and the references therein for more details on the truncated scheme and for a numerical study of the approximation error.  

In the following numerical experiments we choose appropriately the collocation degree $N$, the scaling parameter $\rho$, and the truncation factor $r$, to obtain the desired approximations, specifying our choices each time. 
Typically, satisfactory accuracy can be obtained for $N$ between $20$ and $40$, $r$ between $1/8$ and $1/4$, and $\rho$ chosen so that, for the given integration kernel $K(s),$ the weighted integral is well-defined and accurately approximated by the quadrature rule. In what follows we choose $\mu=2\rho.$
%is the integration kernel and $K(s)e^{\mu s}$ is bounded and vanishes when $s\to+\infty$, then $\mu=2\rho$. 

\section{Numerical study of the emergence of sustained epidemic waves}\label{Sresults}

The simulations shown in the present section aim to illustrate the effectiveness of the numerical method and its applicability to models of the form \eqref{model} that cannot be reduced to ODEs via the LCT. This is the case of gamma-distributed kernels with a non-integer shape parameter. We ran all the simulations using MatCont for MATLAB \cite{MatCont2008}. 

In the following, we take $g(x)=x$ and 
\begin{equation}\label{f1}
    f(x) = \cfrac{\beta_0}{1+kx},\quad x\in \R_{\ge 0},
\end{equation}
where $\beta_0$ is the maximum disease transmission rate, and $k$ is known as the \emph{plasticity/flexibility index} \cite{Zhang2023} or \emph{reactivity factor} \cite{buonomo2023oscillations}: when the incidence is constant at the value $1/k$, the effective contact is halved compared to the baseline in the absence of infections. 
% : $1 + k\times \text{(daily incidence)} = 2$, and then the integral of $K$ is simply the average of a constant function.
With these choices, the EE in \eqref{EEeq} explicitly reads \begin{equation*}
    E_+=\left(\frac{1}{k\gamma} (R_0-1), \ \gamma\right). 
\end{equation*}

As shown in Section~\ref{sec:stability}, it is possible to derive an analytical condition on the parameters determining the stability of the corresponding established equilibrium for certain integer values of~$\alpha$. With these particular choices of $f$ and $g$, 
we have 
\begin{equation*}
    f'(x) = \cfrac{-\beta_0 k}{(1+kx)^2} \quad \text{ and }\quad f^{-1}(y)=\cfrac{1}{k}\left(\cfrac{\beta_0}{y}-1\right),
\end{equation*} 
which, in turn, lead to
$$
C_+=f'(\beta_+ I_+)=\cfrac{-\beta_0 k}{(1+kf^{-1}(\gamma))^2}
%=\cfrac{-\beta_0 k}{\left(1+\cfrac{\beta_0}{\gamma}-1\right)^2}
=-\cfrac{k\gamma^2}{\beta_0}
$$
and 
$$
 \frac{-C_+ \beta_+ I_+}{1-C_+I_+}=\frac{\frac{k\gamma^2}{\beta_0}\beta_+ I_+}{1+\frac{k\gamma^2}{\beta_0}I_+}
 % = \frac{\frac{k\gamma^2}{\beta_0}f^{-1}(\gamma)}{1+\frac{k\gamma}{\beta_0}f^{-1}(\gamma)}
 =\frac{\frac{\gamma^2}{\beta_0}\left(\frac{\beta_0}{\gamma}-1\right)}{1+\frac{\gamma}{\beta_0}\left(\frac{\beta_0}{\gamma}-1\right)}=\frac{\gamma(\beta_0-\gamma)}{2\beta_0-\gamma}.
$$
For $\alpha=2$, using $\tau=2/\mu$, we conclude that the EE is LAS for 
$$ \tau < \frac{4(2\beta_0-\gamma)}{\gamma(\beta_0-\gamma)}. $$
For $\alpha=3$ and $\tau=3/\mu$, the EE is LAS for
\begin{equation*}
    \tau < - \frac{7\beta_0 + \gamma}{3\beta_0} \cdot \frac{2\beta_0-\gamma}{\gamma(\beta_0-\gamma)}. 
\end{equation*}
% \begin{equation*}
%     \mu > - \frac{9}{8+c_0\bar{I}} \frac{c_0\bar{\beta}\bar{I}}{1-c_0\bar{I}} = \frac{9}{7 + \frac{\gamma}{\beta_0}} \frac{\gamma(\beta_0-\gamma)}{2\beta_0-\gamma}
% \end{equation*}
% \ale{In particular, for $\gamma = 1/4$, stability is lost at $ \tau = \cfrac{3\left(7\beta_0+\frac{1}{4}\right)\left(2\beta_0-\frac{1}{4}\right)}{\frac{9}{4}\beta_0\left(\beta_0-\frac{1}{4}\right)}=\cfrac{(28\beta_0+1)(8\beta_0-1)}{3\beta_0(4\beta_0-1)}$}.

% Thus, the stability curve for $\mu = \mu(\beta_0)$ reads
% \begin{align*}
%  \frac{-c_0 \bar{\beta}\bar{I}}{2(1-c_0\bar{I})}=\frac{\frac{k\gamma^2}{\beta_0}\bar{\beta}\bar{I}}{2\left(1+\frac{k\gamma^2}{\beta_0}I\right)}= \frac{\frac{k\gamma^2}{\beta_0}f^{-1}(\gamma)}{2\left(1+\frac{k\gamma}{\beta_0}f^{-1}(\gamma)\right)}=\frac{\frac{\gamma^2}{\beta_0}\left(\frac{\beta_0}{\gamma}-1\right)}{2\left(1+\frac{\gamma}{\beta_0}\left(\frac{\beta_0}{\gamma}-1\right)\right)}=\frac{\gamma(\beta_0-\gamma)}{2(2\beta_0-\gamma)}.
% \end{align*}
% For $\gamma = 1/4$ and $\alpha = 2$, stability is lost at $\tau=\cfrac{16(8\beta_0-1)}{4\beta_0-1}$.

\begin{figure}[ht]
\begin{center}
  \includegraphics[width=\columnwidth]{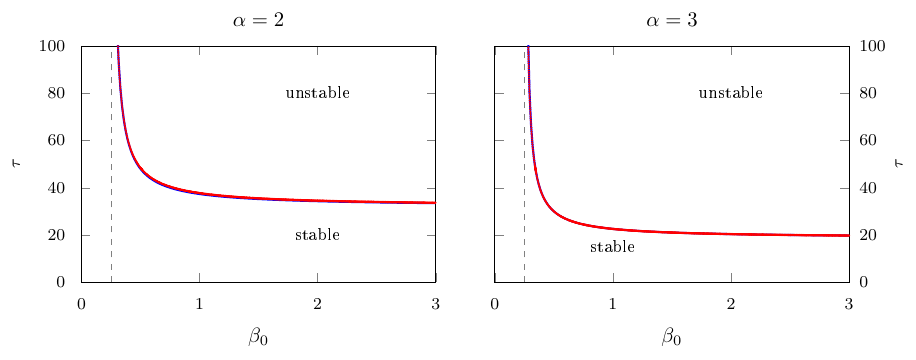}    % The printed column width is 8.4 cm.
\caption{Stable and unstable regions of the EE for \eqref{model} with a gamma-distributed kernel and $f$ defined as in \eqref{f1} in the $(\beta_0,\tau)$ plane, for $\alpha = 2$ (left) and $\alpha = 3$ (right). Blue: analytic curve. Red: obtained with MatCont using $N=20$, $\rho=0.5\alpha/\tau$, and $r=1$. The dashed line delimits the region of existence of the EE (DFE is LAS to the left of the dashed line).
} 
\label{fig:alpha2}
\end{center}
\end{figure}

To validate the numerical method, we first compare the stability curve in the $(\beta_0,\tau)$ plane with the analytic one, for $\alpha = 2,\,3$, where we also fixed $\gamma = 1/4$ (days$^{-1}$) and $k = 0.1$. As shown in Figure \ref{fig:alpha2}, $N=20$ is sufficient for the two curves to almost coincide. 
In the following simulations we normally choose $N=20$, as the curves are usually indistinguishable from those obtained with higher indices, or $N=30$ when the computation of the branches by MatCont requires so. We always tested the simulations with higher values of $N$ (not included here).
%We chose to use the same discretization level for all the subsequent simulations, since the results of our experiments using higher values of $N$ (not included here) are virtually indistinguishable from those with $N=20$.

\begin{figure}[ht]
\begin{center}
\includegraphics[width=\columnwidth]{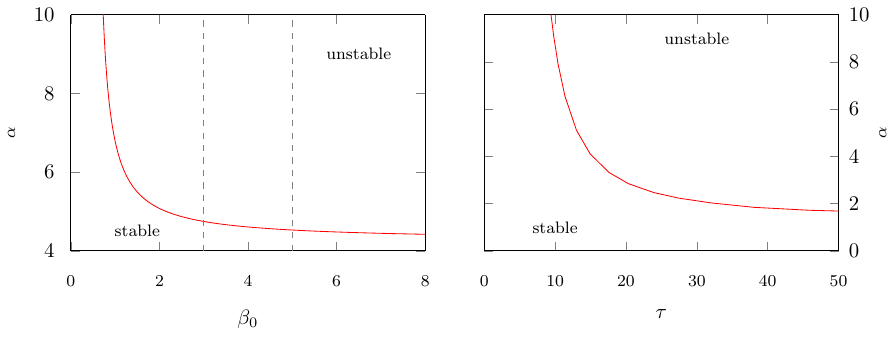}    % The printed column width is 8.4 cm.
\caption{Stable and unstable regions of the EE for \eqref{model} with a gamma-distributed kernel and $f$ defined as in \eqref{f1}. Left: regions in the $(\beta_0,\alpha)$ plane, for $\tau = 14$. Right: regions in the $(\tau,\alpha)$ plane, for $\beta_0=5$. Curves were obtained with MatCont using $N=20$, $\rho=0.5\alpha/\tau$, and $r=1$ (left) and $N=30$, $\rho=0.5\alpha/\tau$, and $r=0.25$ (right). 
} 
\label{fig:tau14}
\end{center}
\end{figure}

Next, we show in Figure \ref{fig:tau14} how our method allows us to numerically compute the stability curves in the $(\beta_0,\alpha)$ plane for a fixed $\tau$, which we set to $\tau = 14$ (days), and in the $(\tau,\alpha)$ plane for a fixed $\beta_0$, which we set to $\beta_0=5$ (days$^{-1}$). Since $\alpha$ varies continuously, the curves cannot have an exact analytic expression, nor can they be obtained numerically using the LCT. 

\begin{figure}[ht]
\begin{center}
  \includegraphics[width=\columnwidth]{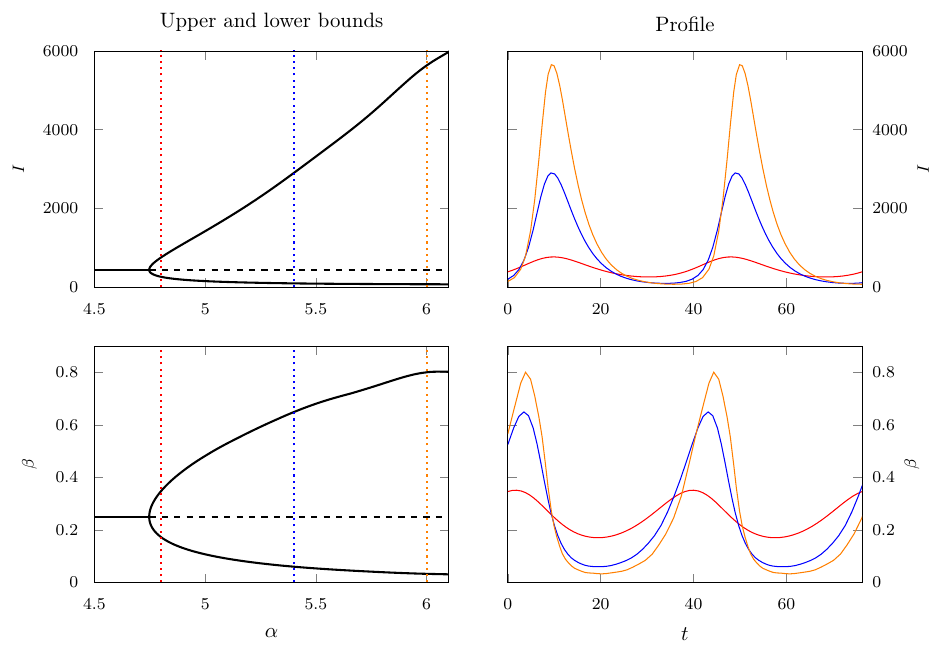}    % The printed column width is 8.4 cm.
\caption{Left: bifurcation diagram for \eqref{model} with a gamma-distributed kernel and $f$ defined as in \eqref{f1} with $\alpha$ as the bifurcation parameter, for $\beta_0=3$ and $\tau = 14$. Right: profiles of the periodic solutions for the values $\alpha = 4.8$ (red), $\alpha = 5.4$ (blue), and $\alpha = 6$ (orange). Curves were obtained with MatCont using $N=20$, $\rho=0.5\alpha/\tau$, and $r=0.25$. 
} 
\label{fig:beta0_3}
\end{center}
\end{figure}

\begin{figure}[ht]
\begin{center}
  \includegraphics[width=\columnwidth]{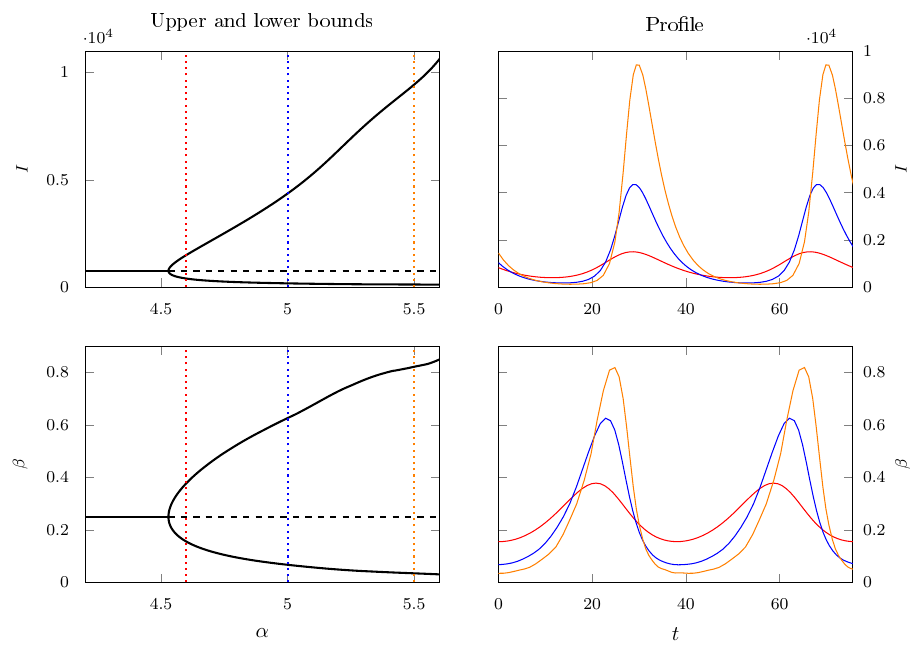}    % The printed column width is 8.4 cm.
\caption{Left: bifurcation diagram for \eqref{model} with a gamma-distributed kernel and $f$ defined as in \eqref{f1} with $\alpha$ as the bifurcation parameter, for $\beta_0=5$ and $\tau = 14$. Right: profiles of the periodic solutions for the values $\alpha = 4.6$ (red), $\alpha = 5$ (blue), and $\alpha = 5.5$ (orange). Curves were obtained with MatCont using $N=20$, $\rho=0.5\alpha/\tau$, and $r=0.25$. 
} 
\label{fig:beta0_5}
\end{center}
\end{figure}

In Figure \ref{fig:tau14} (left), the two values of $\beta_0$  highlighted with dashed gray lines, i.e., $\beta_0=3,\,5$, are those which we use to numerically investigate the one-parameter bifurcation in $\alpha$, as shown in Figures \ref{fig:beta0_3} and \ref{fig:beta0_5}, respectively. In these figures, on the left side, the solid lines in the bifurcation diagrams indicate stable solutions, while the dashed ones represent unstable ones, with maximum and minimum values plotted for the branches of periodic solutions. In other words, the one-parameter continuation shows that the Hopf bifurcation is supercritical. On the right side, we can see three different profiles of periodic solutions corresponding to different values of the parameter $\alpha$. In particular, the solutions in Figure \ref{fig:beta0_3} are obtained with $\beta_0=3$ and $\alpha = 4.8$ (red, period $\approx 38.15$ days), $\alpha = 5.4$ (blue, period $\approx 39.67$ days), and $\alpha = 6$ (orange, period $\approx 40.53$ days). The solutions in Figure \ref{fig:beta0_5} are obtained with $\beta_0=5$ and $\alpha = 4.6$ (red, period $\approx 37.87$ days), $\alpha = 5$ (blue, period $\approx 39.22$ days), and $\alpha = 5.5$ (orange, period $\approx 40.45$ days). By comparing both figures, we can see that the qualitative behavior is very similar, but the amplitudes of the $I$ components vary approximately proportionally with the value of $\beta_0$, while the amplitudes of the $\beta$ components do not seem to be sensitive to variations of that value. Moreover, we can observe that, while the amplitudes of the periodic solutions are rather sensitive to changes in the shape parameter, the effect on the period of oscillations is negligible. 

We conclude this section by considering a more general model where $f$ is defined by 
    \begin{equation}\label{f_mincontact}
        f(x) = \frac{\beta_0+k\beta_1 x}{1+kx}
    \end{equation}
so that $f \to \beta_1 \geq 0$ when $x \to \infty$. Hence, $\beta_1$ represents a minimal transmission rate that can be strictly positive. To the best of the authors' knowledge, this type of response has rarely been taken into account in behavioral modeling. However, from a public health perspective, it is certainly realistic to consider a level of essential activities requiring a minimal amount of daily social contact.

In this case, $f^{-1}(y) = \frac{1}{k} \frac{\beta_0-y}{y-\beta_1}$, so that the EE reads
\begin{equation}\label{EEmincont}
    E_+ = \left(\frac{R_0-1}{k(\gamma-\beta_1)} , \gamma \right).
\end{equation}
Compared to the previous case ($\beta_1=0$), increasing $\beta_1$ leads to a higher number of infections at equilibrium, as expected. 

\begin{figure}[ht]
\begin{center}
  \includegraphics[width=\columnwidth]{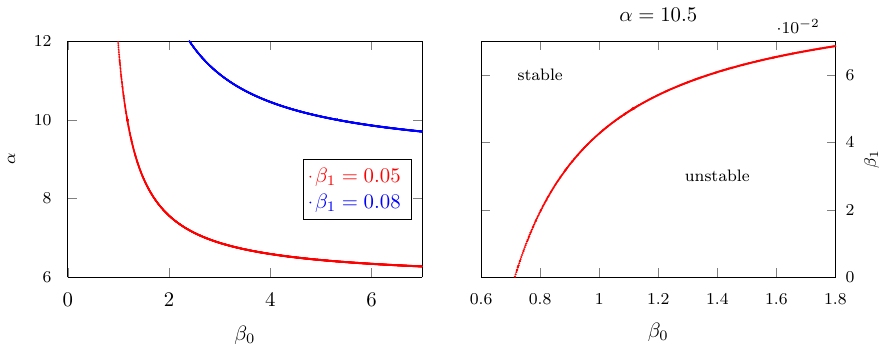}    % The printed column width is 8.4 cm.
\caption{Hopf curves for \eqref{model} with a gamma-distributed kernel and $f$ defined as in \eqref{f_mincontact}, for $\tau = 14$. Left: curves in the $(\beta_0,\alpha)$ plane with $\beta_1 = 0.05$ (red) and $\beta_1 = 0.08$ (blue). In both cases, the stable region is the one below the corresponding curve. Right: curve in the $(\beta_0,\beta_1)$ plane for $\alpha = 10.5$. Curves were obtained with MatCont using $N=20$ and $r=1$. 
} 
\label{fig:mincont_tau14}
\end{center}
\end{figure}

We first fix the additional parameter $\beta_1$ to either $\beta_1=0.05$ or $\beta_1=0.08$ (days$^{-1}$). Figure~\ref{fig:mincont_tau14} (left) shows the sensitivity of the stability curve in the $(\beta_0,\alpha)$ plane to small variations in the value of $\beta_1$. Figure~\ref{fig:mincont_tau14} (right) shows instead the stability curve in the $(\beta_0,\beta_1)$ plane for a fixed non-integer value of $\alpha$ ($\alpha=10.5$). Recall that \eqref{f1} corresponds to \eqref{f_mincontact} for $\beta_1 = 0$. 

We can observe that increasing $\beta_1,$ with $\beta_1<\gamma$, makes the stability region of the EE larger. Even a very small value $\beta_1>0$ is enough to stabilize an equilibrium that would be unstable when $\beta_1=0$. As previously mentioned, a strictly positive $\beta_1$ seems realistic from a public health standpoint.  

\begin{figure}[ht]
\begin{center}
  \includegraphics[width=\columnwidth]{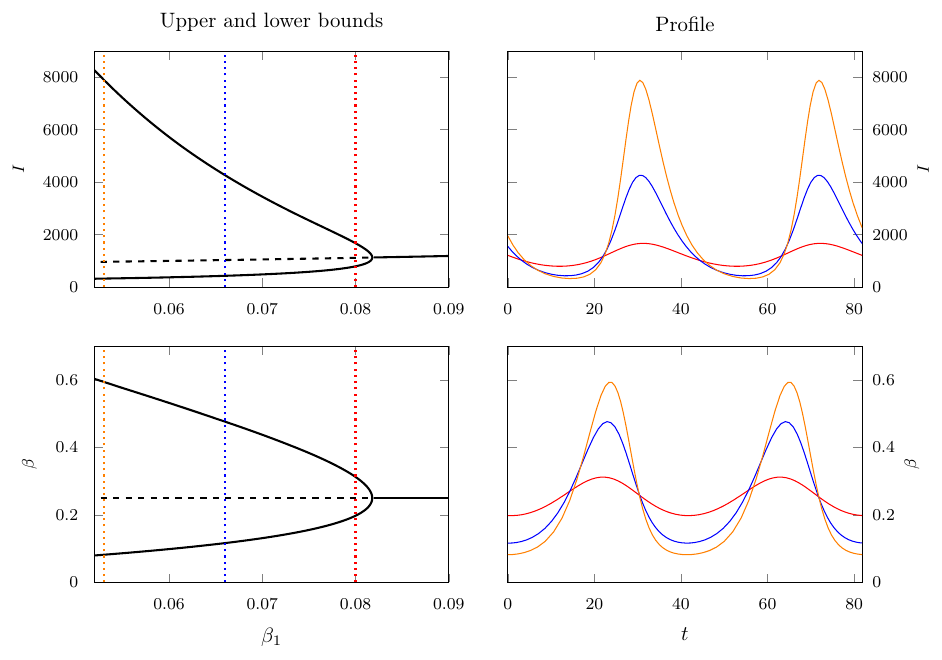}    % The printed column width is 8.4 cm.
\caption{Left: bifurcation diagram for \eqref{model} with a gamma-distributed kernel and $f$ defined as in \eqref{f_mincontact} with $\beta_1$ as the bifurcation parameter, for $\beta_0=5$, $\alpha=10.5,$ and $\tau = 14$. Right: profiles of the periodic solutions for the values $\beta_1 = 0.08$ (red), $\beta_1=0.066$ (blue), and $\beta_1 = 0.053$ (orange). Curves were obtained with MatCont using $N=30$, $\rho=0.5\alpha/\tau$, and $r=0.25$. 
} 
\label{fig:beta1_beta0_5}
\end{center}
\end{figure}

In Figure \ref{fig:beta1_beta0_5} (left), we show the one-parameter bifurcation diagram with respect to $\beta_1$, when $\beta_0=5$. The diagram shows a supercritical Hopf bifurcation, with the branch of periodic solutions existing for small values of $\beta_1$. In the top left picture, we can also see the variation of the number of infections at the equilibrium with respect to $\beta_1$, according to \eqref{EEmincont}. Figure \ref{fig:beta1_beta0_5} (right) shows the profiles of the periodic solutions obtained with $\beta_1=0.08$ (red, period $\approx 40.93$ days), $\beta_1=0.066$ (blue, period $\approx 41.19$ days), and $\beta_1=0.053$ (orange, period $\approx 41.37$ days). Similarly as in the previous analyses, these figures also show that changes in $\beta_1$ can have a substantial impact on the peak of the epidemic waves, but a negligible impact on their period. 

\section{Discussion and concluding remarks}\label{Sconclusion}

In this paper we have applied a recently developed numerical method to extend previous analyses of epidemiological-behavioral models in the literature. 

We have considered an outbreak scenario similar to the one proposed in \cite{Zhang2023}, but with a continuously distributed memory kernel. We confirmed the emergence of oscillations even in the absence of any other process, like demographic turnover or depletion of susceptibles, when the memory kernel is narrow enough around the mean delay. 

In particular, our analyses confirm that sustained oscillations are favored by higher baseline transmission rates and longer delays in the feedback between infection and behavioral response. For the parameter values considered here---representative of a fast-progressing and highly infectious disease---information delays of at least 10--20 days on average appear necessary to generate oscillations. Even in the current era of fast media spread, such long delays are plausible for a fast-progressing disease, and can arise for instance when individuals adjust their behavior in response to severe cases or deaths, see, for instance, \cite{pellisetal}.

Extending previous work of incidence-based behavioral feedback \cite{dOnofrioManfredi2022}, we further investigated the emergence of oscillations as a function of the shape parameter of the memory kernel. Our results show that larger shape parameters, corresponding to a more concentrated distribution, promote oscillatory behavior. In particular, for diseases with low $R_0$ ($R_0\approx 2$) and average information delay of approximately 14 days, oscillations can arise only if the memory kernel is very concentrated around the mean (shape parameter larger than 10). This observation can be intuitively explained by noting that, for a more concentrated kernel, individuals tend to adapt their behavior within a narrower time window, leading to synchronized fluctuations in the transmission rate and, consequently, in infection levels. 

We also examined a more realistic relationship between information and transmission rate by introducing a minimum achievable contact rate representing, for example, essential activities and social interactions. Our analyses indicate that the presence of such a lower bound contributes to stabilize the system at a positive equilibrium, even when the bound is small. Minimal levels of activities have rarely been considered in previous studies, and our findings suggest that they may play an important role on the model dynamics.  
Across all analyses, the amplitude of the oscillations appears to be highly sensitive to parameter variations, whereas the oscillation periods seem to remain relatively stable, with slight variations around 40 days for our parameter values. 

Compared to other previous studies, we were able to extend the analysis to gamma-distributed memory kernels with a general shape parameter by using the pseudospectral approximation method, going beyond the study of Erlang-distributed kernels via the LCT. Using MatCont on the approximating ODE system, we were able to study numerically the branches of periodic solutions emerging from the Hopf bifurcation.
% , giving some insights on the period and amplitude of oscillations when parameters can vary. 
% Compared to other previous studies of incidence-based behavioral feedback \cite{dOnofrioManfredi2022}, we extended the analysis to Gamma-distributed memory kernels with a general shape parameter, going beyond the study of Erlang-distributed kernels via the LCT. Using MatCont, we were able to study numerically the branches of periodic solutions emerging from the Hopf bifurcation, giving some insights on the period and amplitude of oscillations when parameters can vary. 

% We confirmed that oscillations are promoted by: large mean delay, smaller variance around the mean delay (equivalent to larger shape parameter in our tests), large $R_0$. 
% From our tests, the peak of the epidemic waves seem to be rather sensitive to changes in parameters, while the impact on the period seems to be negligible. 

% We also considered a feedback function that accounts for a minimal transmission rate, $\beta_1$, describing essential activities that cannot be suspended. We showed that the incorporation of a minimal level of social contacts has a stabilizing effect on the positive equilibrium. Such minimal level of activities has rarely been considered in previous studies.  

The numerical method is powerful to study general models with distributed memory kernels, but involves the choice of several numerical parameters ($N$, $\rho$, $r$) which require some degree of knowledge of the approximation, and can become computationally expensive when studying dynamics beyond the Hopf bifurcations, like the branches of periodic solutions. 
% The numerical method has been developed for exponentially decaying integration kernels.

\begin{figure}[ht]
\begin{center}
  \includegraphics[width=.8\columnwidth]{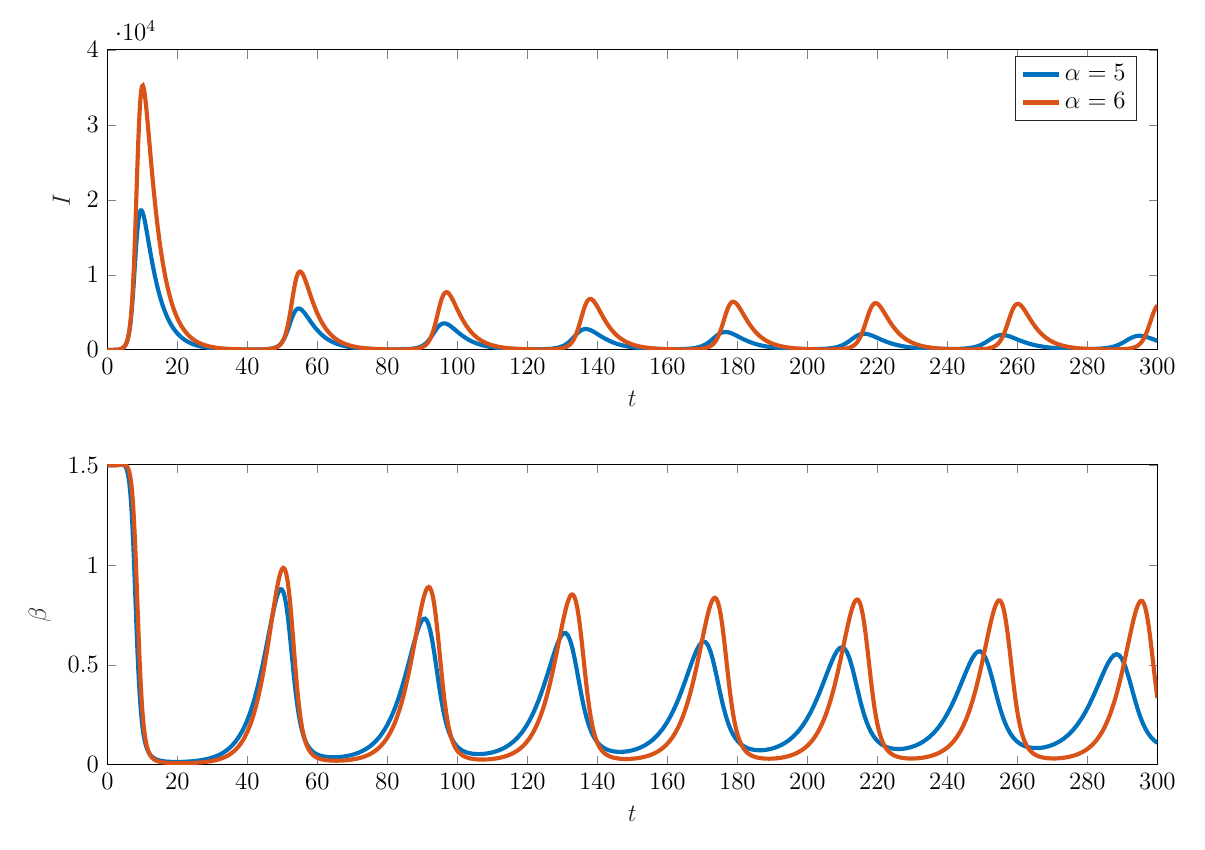}    % The printed column width is 8.4 cm.
\caption{Time integration of the ODE model obtained via LCT. The initial conditions are $I(0)=1$ and the LCT compartments equal to $1/k$. Parameters are $k=0.1$, $\gamma=1/4$, $\beta_0=3$, $\tau=14$. 
Time simulations were obtained using MATLAB's built-in solver \texttt{ode23s} with default error tolerances.}

\label{fig:time_integration}
\end{center}
\end{figure}

While this work focuses on the long-term dynamics of the system, a related interesting question concerns the time scales of emergence of oscillatory dynamics and stabilization toward a limit cycle, which can be investigated through numerical time integration of the original model \eqref{model}. In general the investigation of the dynamics of delay equations by using numerical simulations is challenging as it requires one to select initial functions satisfying suitable compatibility conditions to guarantee that the solutions are sufficiently smooth, see, for instance, \cite{bellen2000} for delay differential equations with finite delay and \cite{brunner2004} for equations of Volterra type. Time integration of the approximating ODE by available ODE software offers an interesting alternative to perform numerical simulation of both delay differential equations and renewal equations with infinite delay. The convergence of the solutions of the approximating ODE to the solutions of the original equation is under investigation. In the case of Erlang-distributed memory kernels, numerical simulations can be performed using the equivalent ODE system obtained via the LCT \cite{dOnofrioManfredi2022}. 
To provide a first insight into the time scales required for reaching the stable attractor in practice, we integrated the LCT-ODE system for two integer values of $\alpha$, for the parameter values in Figure \ref{fig:beta0_3}. The initial condition is $I_0=1$, with all LCT compartments equal to $1/k$, so that the initial information index equals $\beta_0/2$.  
Figure \ref{fig:time_integration} shows that, after an initial large infection wave lasting approximately 30--40 days, the solution rapidly settles on dynamics close to the limit cycle.     
The validity of the low attack rate assumption depends on the total size of the population of interest.

Finally, we mention that an alternative numerical method has been recently proposed to simulate models with distributed memory kernels and infectivity depending on the time since infection \cite{buonomo2025minimal, buonomo2024stable, buonomo2025integral}. These methods allow one to perform time integration of the model preserving the long-term properties of the system and, to our understanding, can also be applied to general memory kernels $K$. We also refer to \cite{andovermiglio,andovermiglioIFAC} for time-integration methods of finite delay equations based on exponential integrators. While all these methods can tackle integral equation models, the pseudospectral approach has the advantage to enable users to apply established software tools for ODEs, including their graphical user interface (like in the case of MatCont). 

\appendix

\section{Computations for the stability analysis of equilibria}\label{CompLinStab}
In this section, we provide detailed computations for the results presented in \Cref{sec:stability} on the linear stability of the EE $E_+$ in \eqref{EEeq}.
We assume that $K$ is a gamma probability density function with shape parameter $\alpha$ and rate parameter $\mu$. Then, the Laplace transform of $K$ reads as in \eqref{KLaplace}, and the associated characteristic function $\Delta$ reads as in \eqref{quasipol}.
Note that, if $\alpha\to \infty$ in such a way that $\alpha/\mu \to \tau < \infty$, then $K$ tends in distribution to a Dirac delta concentrated at $\tau$.
If $\alpha$ is an integer, then $K$ is an Erlang probability density function, and $\Delta$ is a polynomial. Hence, one can analyze the stability of $E_+$ with the Routh--Hurwitz criterion \cite{dOnofrioManfredi2022}.
Here we derive conditions for the LAS of the EE in three different cases, namely $\alpha=1, 2, 3$ and $\mu>0$.
\begin{itemize}
\item Case $\alpha=1$  (exponentially fading memory kernel): The Laplace transform of $K$ is $$\widehat{K}(\lambda)=\frac{\mu}{\mu+\lambda},$$ and the characteristic equation \eqref{CE} reads 
\begin{equation*}
 \lambda - \frac{\mu C_+ I_+ \lambda }{\mu+\lambda} - \frac{ \mu C_+ \beta_+I_+ }{\mu +\lambda} = 0.
\end{equation*}
We look for solutions such that $\Re\lambda > -\mu$.  To do this, we are led to solve the quadratic equation 
\begin{equation*}
    \lambda^2 + \mu \lambda (1 - C_+ I_+ ) - \mu C_+ \beta_+ I_+ = 0,
\end{equation*}
whose solutions are given by
\begin{equation*}
    \lambda_{1,2} = \frac{-\mu (1 - C_+I_+ ) \pm \sqrt{ \mu^2 (1 - C_+ I_+ )^2 + 4 \mu C_+ I_+ \beta_+} }{2} \, .
\end{equation*}
Since $C_+<0$, the term under the square root is nonnegative if 
\begin{equation*}
    \mu \geq \frac{-C_+\beta_+I_+}{(1-C_+I_+)^2},
\end{equation*}
and in this case both roots are real and negative. In the other case, the characteristic roots are complex conjugate, with a negative real part. Hence, the EE is LAS for all $\mu>0$, which is in agreement with what was observed in \cite{dOnofrioManfredi2022} for the case of models  including depletion of susceptibles and demography.
\item Case $\alpha=2$: The characteristic equation \eqref{CE} reads 
\begin{equation*}
    \lambda^3 + 2\mu\lambda^2 + \mu^2 (1-C_+I_+) \lambda - C_+ \mu^2 \beta_+I_+ = 0. 
\end{equation*}
The Routh--Hurwitz criterion for a third-order polynomial $a_3\lambda^3+a_2\lambda^2+a_1\lambda+a_0=0$ states that the EE is LAS if all coefficients of the characteristic function are strictly positive and $a_1a_2>a_0a_3$. In this case, the condition becomes 
\begin{equation*}
    2\mu(1-C_+I_+) > -C_+ \beta_+I_+ \quad \Leftrightarrow \quad \mu > \frac{-C_+ \beta_+I_+}{2(1-C_+I_+)}.
\end{equation*}
Thus, $E_+$ can destabilize already for $\alpha=2$ if $\mu$ is small enough (or, equivalently, if the delay $\tau=2/\mu$ is large enough).  
\item Case $\alpha=3$: The characteristic equation \eqref{CE} reads 
\begin{equation*}
    \lambda (\mu+\lambda)^3 - C_+ \mu^3 I_+ \lambda -C_+ \mu^3 \beta_+I_+ = 0. 
%    \quad \Leftrightarrow \quad \lambda^3 + 2\mu\lambda^2 + \mu^2 (1-c_0\bar{I}) \lambda - c_0 \mu^2 \bar{\beta}\bar{I} = 0. 
\end{equation*}
The Routh--Hurwitz criterion for a fourth-order polynomial $a_4\lambda^4+a_3\lambda^3+a_2\lambda^2+a_1\lambda+a_0=0$ states that the EE is LAS if all coefficients of the characteristic function are strictly positive, and $a_1a_2>a_0a_3$, and $a_3a_2a_1>a_4a_1^2+a_3^2a_0$. 

In this case $a_4=1$, $a_3=3\mu$, $a_2=3\mu^2$, $a_1=\mu^3-C_+\mu^3I_+$, $a_0= - C_+ \mu^3 \beta_+I_+$. Hence, the conditions for the EE to be LAS are 
\begin{align*}
    & 3\mu^5(1-C_+I_+) > -3 C_+ \mu^4\beta_+I_+ \quad \Leftrightarrow \quad \mu > \frac{- C_+I_+\beta_+}{1-C_+I_+} 
\end{align*}
and
\begin{align*}
    & 9\mu^6(1-C_+I_+) > \mu^6 (1-C_+I_+)^2 - 9C_+ \mu^5  \beta_+ I_+ 
    \quad \Leftrightarrow \quad \left(\frac{8+C_+I_+}{9} \right) \mu > -\frac{C_+\beta_+I_+}{1-C_+I_+}. 
\end{align*}
\end{itemize}

\section*{Acknowledgements} 
The authors thank Piero Manfredi for his valuable comments which have contributed to improving the final version of the manuscript.
This project started within a research visit supported by the University of Leeds Research Equity, Diversity and Inclusion (REDI) Pilot Fund and the School of Mathematics Research Visitor Centre. 
The work of AA, SDR, and RV  was partially supported by the Italian Ministry of University and Research (MUR) through the PRIN 2020 project (No.\ 2020JLWP23) ``Integrated Mathematical Approaches to Socio-Epidemiological Dynamics'', Unit of Udine (CUP: G25F22000430006). SDR was supported by the project ``One Health Basic and Translational Actions Addressing
Unmet Needs on Emerging Infectious Diseases'' (INF-ACT), BaC ``Behaviour and sentiment monitoring and modelling for outbreak control/BEHAVE-MOD'' (No.\ PE00000007, CUP I83C22001810007)
funded by the NextGenerationEU.
AA, SDR, FS, and RV are members of the INdAM research group GNCS and UMI research group Mo\-del\-li\-sti\-ca Socio-Epidemiologica. FS is a member of JUNIPER (Joint UNIversities Pandemic and Epidemiological Research). 

%\section*{Authors contribution}

\bibliographystyle{aims} 
\providecommand{\href}[2]{#2}
\providecommand{\arxiv}[1]{\href{http://arxiv.org/abs/#1}{arXiv:#1}}
\providecommand{\url}[1]{\texttt{#1}}
\providecommand{\urlprefix}{URL }

\end{document}